\documentclass[aps,prmaterials,twocolumn,superscriptaddress]{revtex4-2}
\pdfoutput=1
\usepackage{graphicx}
\graphicspath{{./figures/}}
\usepackage[pdftex, pdftitle={Article}, pdfauthor={Author},hidelinks]{hyperref} 

\newcommand{\Ic}{$I_c$}
\newcommand{\Jc}{$J_c$}
\newcommand{\Tc}{$T_c$}
\newcommand{\Hirr}{$H_{irr}$}
\newcommand{\Hk}{$H_k$}
\newcommand{\BSSCO}{Bi$_2$Sr$_2$Ca$_1$Cu$_2$O$_x$}
\newcommand{\degree}{$^{\circ}$}

\begin{document}

\title{The Conundrum of Strongly Coupled Supercurrent Flow in Both Under- and Over-doped Bi-2212 Round Wires}
\author{Yavuz Oz}
\author{Jianyi Jiang}
\affiliation{Applied Superconductivity Center, National High Magnetic Field Laboratory, Florida State University, 2031 East Paul Dirac Drive, Tallahassee, Florida, 32310, USA}
\author{Maxime Matras}
\author{Temidayo Abiola Oloye}
\author{Fumitake Kametani}
\author{Eric E. Hellstrom}
\author{David C. Larbalestier}
\email[Correspondence and requests should be addressed to David Larbalestier: ]{larbalestier@asc.magnet.fsu.edu}
\affiliation{Applied Superconductivity Center, National High Magnetic Field Laboratory, Florida State University, 2031 East Paul Dirac Drive, Tallahassee, Florida, 32310, USA}
\affiliation{FAMU-FSU College of Engineering, Tallahassee, Florida, 32310, USA}
\date{\today}

\begin{abstract}
Understanding what makes \BSSCO{} (Bi-2212) the only high critical current density (\Jc{}), high temperature superconductor (HTS) capable of being made as a round wire (RW) is important intellectually because high \Jc{} RW Bi-2212 breaks the paradigm that forces biaxially textured REBCO and uniaxially textured (Bi,Pb)$_2$Sr$_2$Ca$_2$Cu$_3$O$_x$ (Bi-2223) into tape geometries that reproduce the strong anisotropy of the native crystal structure and force expensive fabrication routes to ensure the best possible texture with minimum density of high angle grain boundaries. The \textbf{biaxial} growth texture of Bi-2212 developed during a partial melt heat treatment should favor high \Jc{}, even though its $\sim$15\degree{} full width at half maximum (FWHM) grain-to-grain misorientation is well beyond the commonly accepted strong-coupling range. Its ability to be strongly overdoped should be valuable too, since underdoped cuprate grain boundaries are widely believed to be weakly linked. Accordingly, we here study property changes after oxygen underdoping the optimized, overdoped wire. While \Jc{} and vortex pinning diminish significantly in underdoped wires, we were not able to develop the prominent weak-link signature (a hysteretic \Jc{}(H) characteristic) evident in even the very best Bi-2223 tapes with a $\sim$ 15\degree{} FWHM \textbf{uniaxial} texture. We attribute the high \Jc{} and lack of weak link signature in our Bi-2212 round wires to the high-aspect ratio, large-grain, basal-plane-faced grain morphology produced by partial-melt processing of Bi-2212 which enables \textit{c}-axis Brick-Wall current flow when \textit{ab}-plane transport is blocked. Since we have elsewhere concluded that only a very small fraction of the filament cross-section carries the current, we conclude that the presently optimized biaxial texture of Bi-2212 intrinsically constitutes a strongly coupled current path, regardless of its oxygen doping state.
\end{abstract}


\maketitle

\section{Introduction}

\begin{figure*}
	\includegraphics[width=2.0\columnwidth]{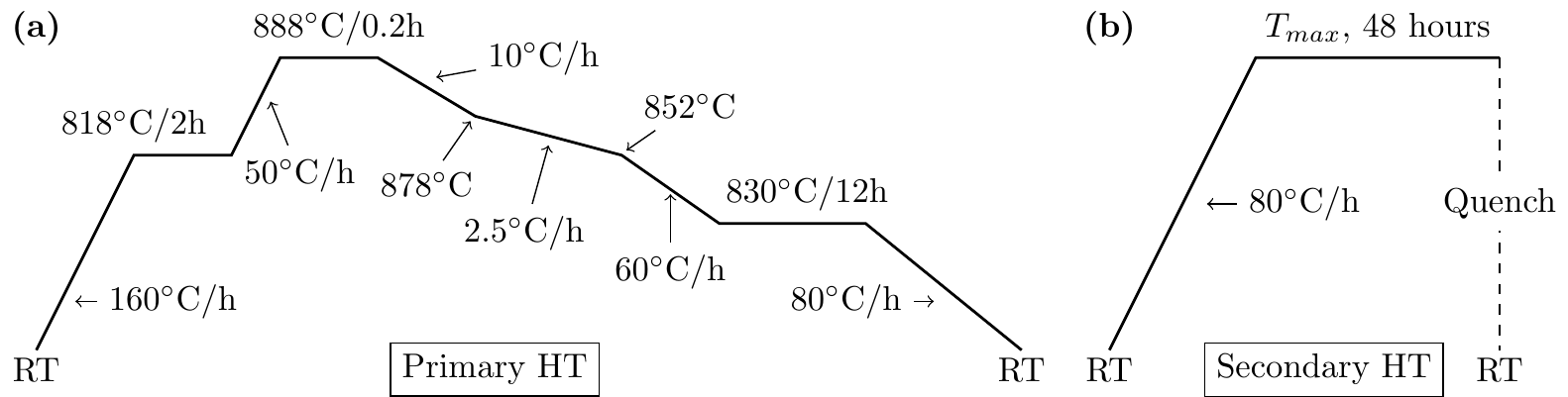}  
	\caption{\label{fig:HT} The primary (a), and secondary (b) heat treatment procedures used to process the samples. Primary HT was carried out at 100 bar, whereas the secondary HT was carried out at 1 bar.}
\end{figure*}

Randomly oriented grain boundaries (GBs) are known to be a major impediment to the realization of high \Jc{} in polycrystalline HTS \cite{Hilgenkamp2002Grain,Larbalestier2014Isotropic}. The most widely studied and clearly problematic GBs are planar [001] tilt GBs \cite{Hilgenkamp2002Grain}. Dimos \textit{et al.} \cite{Dimos1988Orientation,Dimos1990Superconducting} showed early on that \Jc{} across planar GBs decreases exponentially with increasing grain-to-grain planar [001] tilt misorientation angle in YBCO, with a critical angle for the onset of weak link behavior later shown to be as low as 2\degree{}- 3\degree{} \cite{verebelyi,feldmann}. [001] tilt GBs in Bi-2212 \cite{amrein} and Bi-2223 \cite{attenberger,hanisch} display a similar dependence. Thus, the high \Jc{} YBCO and Bi-2223 anisotropic conductor forms needed for applications are made by complex and expensive fabrication processes to achieve the desired texture that minimizes the high angle grain boundary (HAGB) density \cite{Larbalestier2014Isotropic,Kametani2015Comparison}. In marked contrast, recent work has developed Bi-2212 into a high \Jc{} round wire conductor \cite{Larbalestier2014Isotropic,Kametani2015Comparison,Shen2009Development} without weak-link character and with similar architecture to low temperature superconductor (LTS) wires. The round-wire form is very desirable for magnet applications because it allows the conductor to be produced in a low-hysteretic-loss, multi-filamentary, twisted state in a macroscopically isotropic architecture that provides high magnetic field quality. The paradox of round wire Bi-2212 is that it is now possible to make it with some 6 times higher \Jc{} \cite{shen2019stable,brown,Jiang2019High} than carefully textured Bi-2223 tapes \cite{kametani2019visualization,sato}, a conundrum that seems to violate much of our hard-won understanding of HTS grain boundaries over the last 30 years.

A key step that enabled high \Jc{} in Bi-2212 was to fully densify filaments using an overpressure heat treatment, providing nearly perfect physical connectivity \cite{Larbalestier2014Isotropic}. However, full densification does not provide a GB network without blocking misorientations, or answer the question of how Bi-2212 is able to deliver much higher \Jc{} compared to uniaxially textured Bi-2223 tapes, where a lengthy mechanical and thermal processing development only slowly led to practical \Jc{} values \cite{sato,rogalla}. One part of the puzzle is provided by a recent comparison \cite{Kametani2015Comparison} of texture in high \Jc{} Bi-2223 and Bi-2212, which showed that a quasi-biaxial texture with a full width at half maximum (FWHM) of $\sim$15\degree{} can be developed by first melting the Bi-2212 inside the $\sim$15 $\mu$m diameter filaments and then slowly solidifying under sparse nucleation conditions that optimize rapid growth of long \textit{a}-axis textured grains along the filament axis. Surprisingly, despite a 15\degree{} FWHM being well into the weak link regime of [001] tilt YBCO bicrystals, round wire Bi-2212 can achieve a transport \Jc{} many times that of the state of the art $\sim$15\degree{} FWHM uniaxially textured Bi-2223 tape \cite{kametani2019visualization}. Moreover, the Bi-2212 wires lack the hysteretic field-increasing, field-decreasing, weak-link signature displayed by Bi-2223 \cite{Kametani2015Comparison}. Indeed a recent study found evidence of well-connected grains in an untextured bulk Bi-2212 sample but not in a similarly prepared Bi-2223 bulk sample \cite{wang2018cause}. This paradox of greatly divergent connectivity between two very similar siblings is what has stimulated the present study.  We know that round wire Bi-2212 shows great promise for ultra high field solenoids \cite{Larbalestier2014Isotropic, nationalresearch}, for plasma fusion use \cite{liu2018experimental} and for accelerator dipoles \cite{shen2019stable}.  All such applications are made more attractive by higher \Jc{} values and because we conclude that the \Jc{} of the best present Bi-2212 \cite{brown, Jiang2019High} is still less than 1\% of Bi-2212’s depairing current density \cite{wang2018cause, Shen2010Filament}, it seems clear that the active superconducting cross section is significantly less than unity. Understanding where obstacles to current flow occur and what they are is key to further enhancing the \Jc{}. We note that a recent change of Bi-2212 powder type to one of more uniform and finer particle size has enabled a doubling of the maximum \Jc{} compared to the already good powder made by Nexans used in the wires of this present study \cite{shen2019stable,brown, Jiang2019High}.

All models of current flow in HTS compounds emphasize the primacy of \textit{ab}-plane current flow, only considering other possible current paths when this path is blocked. Two prominent models for current transport emerged from early studies of the grain morphology of Bi-2223 tapes. The Brick-wall (BW) model \cite{Bulaevskii1992Model,Bulaevskii1993Limits} proposed that [001] \textit{c}-axis twist GBs provide a strongly-linked \textit{c}-axis supercurrent path when the \textit{ab}-plane was blocked, as a result of the large basal-plane-faced GB surface area provided by high aspect ratio grains. In contrast, the Railway-switch (RS) model \cite{Hensel1993model,Hensel1995Limits} proposed that the \textit{ab}-plane supercurrent path connects largely by low-angle, obliquely intersecting strongly coupled GBs called small-angle, \textit{c}-axis tilt (SCTILT) grain boundaries. Steadily improving texture was then thought to minimize the need to access the intrinsically lower \Jc{} \textit{c}-axis paths between the dominant \textit{ab}-plane paths, allowing an increase in overall \Jc{} of Bi-2223 tapes by better occupancy of the tape cross-section by the transport current. Note that both models predict intra- and inter-grain \textit{c}-axis transport to be a factor limiting the macroscopic \Jc{}. The much lower magnitude of \textit{c}-axis \Jc{} is at least partially compensated by the large area of the basal-plane-faced \textit{c}-axis contacts in the BW case. Indeed, an earlier study by Cai \textit{et al.} \cite{Cai} found direct evidence for \textit{c}-axis current paths in high-\Jc{} micro-sections of individual Bi-2223 filaments. In the context of the Brick Wall and Railway Switch models, Kametani \textit{et al.} \cite{Kametani2015Comparison} offered possible explanations for the high \Jc{} and scarcity of weak links in Bi-2212 round wires, despite their comparatively high 15\degree{} FWHM GB misorientation, by focusing on particular differences of the Bi-2223 and Bi-2212 grain morphology. Although both compound wires share a colony grain structure of stacked grains with a common \textit{c}-axis, large [001] twist misorientations are much more common in Bi-2223 than in Bi-2212, as are \textit{c}-axis tilts. Bi-2212 has a much more ordered colony structure arising from the sparse grain nucleation and rapid growth of \textit{a}-axis-oriented grains along the filament axis, which consume slower-growing, less favorably oriented grains on cooling from the liquid state. The result is a strong [100] \textit{a}-axis texture with [010] at right angles and a slowly rotating [001] direction along the filament.  In a conductor with several hundred such filaments, the overall effect is to produce a macroscopically isotropic behavior, even though each grain retains its full intrinsic electronic anisotropy.  Because this Bi-2212 grain structure develops slowly from the melt, a much higher aspect ratio brick wall with much larger grains can form than in Bi-2223 where only a marginal liquid is present and the reaction to form Bi-2223 occurs largely in the solid state.  An additional non-trivial factor favoring higher \Jc{} in Bi-2212 is that, unlike Bi-2223 and YBCO which can only be marginally overdoped, Bi-2212 can be strongly overdoped \cite{Shen2009Development,li_strongly_underdoped_single_crystals}, potentially improving its GB connectivity, since GBs in cuprates are widely believed to be underdoped with respect to the grains \cite{Shen2009Development,Li2015Study}. Overdoping also enhances intra-grain vortex pinning by reducing electromagnetic anisotropy $\gamma$ \cite{Shen2009Development,li_strongly_underdoped_single_crystals,Nakane2004Performance,Nakane2004Effect,Shimoyama1997Strong,Kishio1994Carrier,Nakane2005Effect,Kumakura1996Effect,Haraguchi2005Critical,Correa2001Overdoping,Piriou2008Effect,Fukumoto1996Effect}, which improves vortex stiffness, helps grain-to-grain coupling \cite{Kumakura1996Effect,Fukumoto1996Effect}, and increases the irreversibility field ($H_{irr}$), all of which together enhance \Jc{} \cite{Shen2009Development,Nakane2004Performance,Nakane2004Effect,Shimoyama1997Strong,Kishio1994Carrier,Nakane2005Effect,Kumakura1996Effect,Haraguchi2005Critical}. In summary one can clearly see many similarities between Bi-2223 and Bi-2212, but what matters today is to clearly distinguish the important differences that actually make 6 times higher overall \Jc{} possible in Bi-2212, whether by better connectivity or by better vortex pinning or some combination of the two. Technologically, it is a great advantage of Bi-2212 that very high \Jc{} is now possible in the desirable round-wire, isotropic architecture preferred for conductor use, especially one with great architectural flexibility with respect to filament number, arrangement and size.

Here we focus on one key attribute of Bi-2212, namely its very broad doping range.  Presently optimized Bi-2212 is significantly overdoped, which is expected to compensate for any GB oxygen deficiency compared to the grains. With the expectation that global underdoping treatments would  remove this compensation, we here investigate how optimized high \Jc{}, oxygen-overdoped Bi-2212 round wires behave when systematically underdoped so as to better understand the extent to which oxygen doping affects GB connectivity, vortex pinning and transport \Jc{}. Previous studies on Bi-2212 have reported increased \Hirr{} in Bi-2212 tapes \cite{Nakane2004Performance,Kumakura1996Effect}, single crystals \cite{Shimoyama1997Strong,Kishio1994Carrier,Haraguchi2005Critical}, and round wires \cite{Shen2009Development,Pallecchi2017Investigation}, improved \Jc $(H)$ \cite{Shen2009Development,Nakane2004Performance,Nakane2004Effect,Shimoyama1997Strong,Nakane2005Effect,Kumakura1996Effect,Haraguchi2005Critical}, and changes in critical temperature (\Tc{}) \cite{Shen2009Development,li_strongly_underdoped_single_crystals,Nakane2004Performance,Nakane2004Effect,Nakane2005Effect,Haraguchi2005Critical,Correa2001Overdoping,Piriou2008Effect,Yamashita2010Control} with increasing oxygen doping level. For the present study, as illustrated in Figure \ref{fig:HT}, optimally overdoped (with respect to \Jc{}) and full-density, overpressure-processed Bi-2212 round wires originally processed at 100 bar total pressure (with 1 bar O$_2$, balance Ar) were each subjected to a secondary heat treatment (HT) at a low oxygen partial pressure (pO$_2$) to progressively underdope them, with the expectation that higher secondary heat treatment temperature would result in lower oxygen content in both grains and grain boundaries.

\begin{figure*}
	\includegraphics[width=1.9\columnwidth]{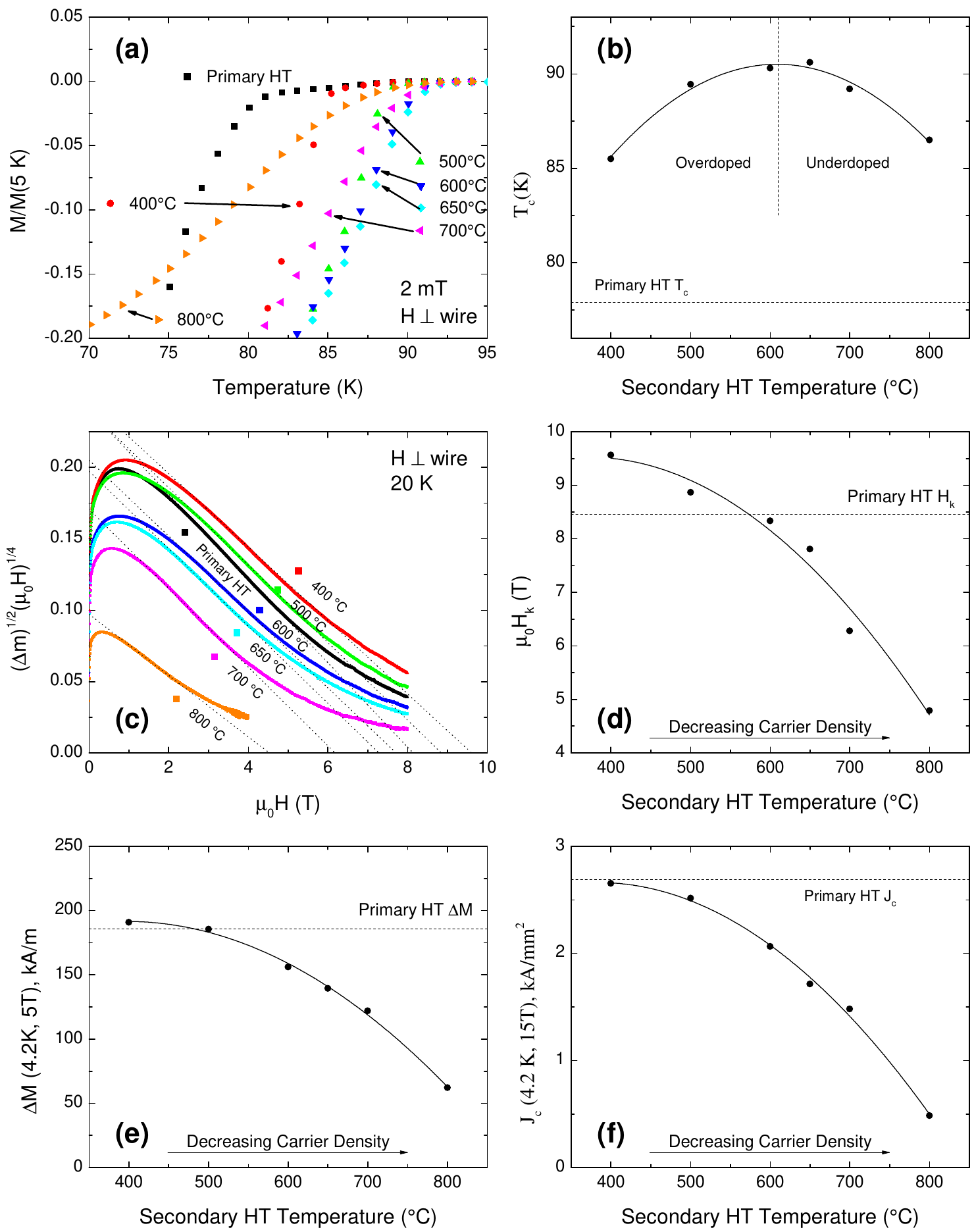}
	\caption{\label{fig:main} a) Zero field cooled (4.2 K) magnetization, measured in a SQUID magnetometer with H = 2 mT applied perpendicular to the wire axis. As is typical for largely decoupled Bi-2212 filaments, the transitions are all broad. b) \Tc{} versus secondary heat treatment temperature ($T_{max}$). As $T_{max}$ increased, the samples transitioned from the initially overdoped state to the underdoped state, reducing the carrier density. c) Kramer function calculated from $\Delta m$ measurements. The linear regime after the initial peak can be extrapolated to zero to obtain the Kramer field ($H_k$). d) Kramer field, a measure of \Hirr{}, at 20 K versus $T_{max}$. Vortex pinning, as judged by \Hirr{}, decreased with the decreasing carrier density induced by increasing $T_{max}$. e) Magnetization hysteresis ($\Delta M$ at 5 T, 4.2 K) versus $T_{max}$. $\Delta M$ decreased as the samples were increasingly underdoped, much like transport \Jc{}. f) Transport \Jc{} at 15 T,4.2 K versus $T_{max}$. Consistent with $\Delta M$ (e), transport \Jc{} went down as charge carriers were depleted. The horizontal dashed lines denote the value for the sample that underwent only the primary heat treatment. The curved lines are guides for the eye.}
\end{figure*}
\section{Methodology}
\subsection{Sample Preparation}
The Ag-Mg sheathed multifilamentary Bi-2212 round wires contained 18 bundles of 37 filaments (37x18), with as-drawn filament and wire diameters of $\sim$15 $\mu$m/0.8 mm. They were fabricated by Oxford Superconducting Technology (OST) using the Powder-In-Tube technique with Nexans granulate Bi-2212 powder manufactured in 2013. Although state-of-the-art at the time, this powder has since been substantially bettered by finer and more uniform powder manufactured by Engi-Mat that allows twice higher \Jc{} \cite{Jiang2019High}. Approximately 1.2 m long pieces were melt processed under 100 bar (1 bar O$_2$, balance Ar) using the heat treatment profile of Figure \ref{fig:HT}a and small pieces were then underdoped using a secondary heat treatment shown in Figure \ref{fig:HT}b, in a flowing gas, quench furnace under a 1 bar mixture of Ar-O$_2$ with a pO$_2$ of 0.01 bar. The maximum temperature ($T_{max}$) was varied between 400 \degree{}C and 800 \degree{}C to change the doping state, with increasing $T_{max}$ values leading to lower oxygen content. The furnace was ramped at a constant rate of 80 \degree{}C per hour from room temperature ($\sim$30 \degree{}C) to $T_{max}$ and equilibrated at $T_{max}$ for 48 hours, after which the sample was pulled out of the hot zone to lock in the doping state.  

Samples were prepared for Electron Backscatter Diffraction Orientation Imaging Microscopy (EBSD-OIM) by polishing with diamond films of decreasing grit sizes, followed by ion polishing in a JEOL Cross Section Polisher using an Argon (Ar) beam at 6 kV, and finally surface-cleaning at 2 kV.

\subsection{Measurements and Data Analysis}
Standard 4-point transport measurements of the critical current \Ic{} were conducted at field values ranging from 1 to 15 Tesla (T) in a superconducting magnet using a criterion of 1 $\mu$V/cm. The field, applied perpendicular to the direction of current flow, was cycled in 1 T increments from 1 T to 15 T and then back to 1 T to check for hysteresis in the \Ic{}(H) curves, \Ic{} being measured at each field step. Conversion to \Jc{} was made using the fully dense Bi-2212 areas measured by optical image processing after a densification heat treatment at 820 \degree{}C \cite{matras2020process}. Filaments were observed to be 60-70\% dense before reaction and greater than 95\% dense after melt processing.

To compare the doping state of the samples, \Tc{} values were derived from measurements of magnetic moment as a function of temperature. The samples were zero field cooled to 4.2 K in a SQUID magnetometer, and then warmed up to 100 K at 0.25 K/minute in a 2 mT field applied perpendicular to the wire axis. \Tc{} was defined as the linear extrapolation of the $m$(H) data to $m$ = 0 just before the sample fully transitioned to the normal state.

Magnetization hysteresis measurements were made at 4.2 K and 20 K in a 14 T Vibrating Sample Magnetometer (VSM) with the wire axis perpendicular to the field. The magnetization loops $m$(H) were run from -2 T to 14 T to -2 T at a constant rate of 10 mT/s and magnetization hysteresis ($\Delta M$) values were extracted, where M is the sample magnetic moment normalized to the superconducting Bi-2212 volume and $\Delta M$ is defined as $M$(field decreasing) – $M$(field increasing). The Kramer field (\Hk{}) values were used to estimate \Hirr{}, the field at which \Jc{} drops to zero, by linear extrapolation of the $(\mu_0 H)^{1/4} (\Delta m)^{1/2}$ versus field data at 20K to the H axis to define \Hk{} \cite{kramer}.
\section{Findings}
Figure \ref{fig:main}a shows changes in the \Tc{} transitions as a function of the secondary heat treatment $T_{max}$. We note that the initial onset \Tc{} is below 80 K, clearly overdoped and \Tc{} rises steadily to over 90 K with underdoping, before dropping again as the underdoping increases. The broad magnetization transitions are typical of Bi-2212 where the filaments are electromagnetically decoupled from each other and the filament diameter is small. Extrapolating the high temperature portion of the curve to zero magnetization allows us to define an upper limit \Tc{}, shown in Figure \ref{fig:main}b.  \Tc{} clearly follows the parabolic trend typical of the relationship between \Tc{} and oxygen/carrier content in Bi-2212 \cite{Piriou2008Effect,Yamashita2010Control}, in agreement with the expectation that increasing $T_{max}$ leads to a continuous decrease in oxygen content. The primary sample, which was cooled to ambient temperature in 1 bar O$_2$, was the most overdoped. When $T_{max}$ was between 400 \degree{}C and 600 \degree{}C, \Tc{} rose significantly from $\sim$86 K to $\sim$90 K. Around $T_{max}$ = 800 \degree{}C the transition became very extended, suggesting that cation composition changes, not just oxygen doping.

Figure \ref{fig:main}c shows Kramer function plots taken at 20 K with their linear extrapolation \Hk(20 K) values given in Figure \ref{fig:main}d. \Hk(20 K) is initially improved by the 400 \degree{}C secondary HT compared to the primary HT, rising from 8.5 T to 9.5 T. However, after this initial increase, \Hk{} decreases monotonically with increasing secondary HT temperature. Using \Hk{} to approximate \Hirr{}, we interpret this decrease as a continuous decline in vortex pinning strength with an increasing electronic anisotropy induced by the decreasing carrier density \cite{Nakane2004Performance,Correa2001Overdoping}.

Like \Hk{}, the magnetization hysteresis $\Delta M$(4.2 K, 5 T) (Figure \ref{fig:main}e) for the 400 \degree{}C secondary HT samples showed a small increase above the primary HT samples, but quickly declined with increasing secondary HT temperature. The trend is similar to that of transport \Jc{}, as magnetization is proportional to the magnitude of currents circulating in the wire and the lengths over which these currents circulate \cite{angadi}.

\begin{figure*}
	\includegraphics[width=1.5\columnwidth]{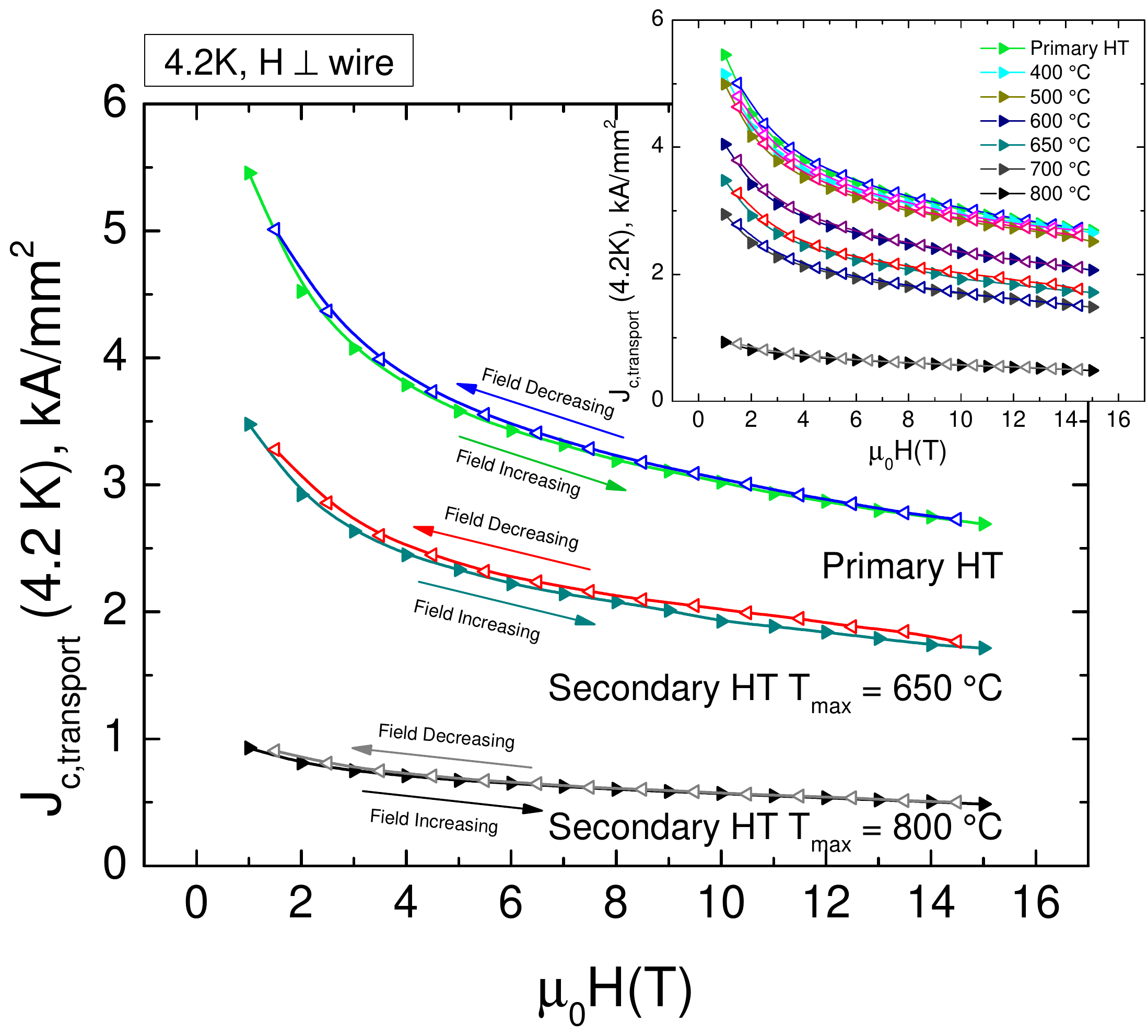}  
	\caption{\label{fig:hysteresis} Transport \Jc{}(4.2 K) in increasing and decreasing perpendicular applied field for the samples secondary annealed at 650 \degree{}C, 800 \degree{}C compared to the primary heat treatment sample. The data for all samples are shown in the inset. \Jc{}(H) for samples secondary annealed at $T_{max}$ = 600 \degree{}C and higher are significantly smaller than the optimally overdoped sample, yet none show any significant hysteresis.}
\end{figure*}

Figure \ref{fig:main}f plots transport \Jc{}(4.2 K, 15 T), the most important performance metric for a superconducting wire. Transport \Jc{} follows a trend like that of $\Delta M$(4.2 K, 5 T), decreasing with increasing secondary annealing temperature. However, unlike \Hk{} and $\Delta M$, transport \Jc{} was not initially enhanced by a secondary HT, perhaps because the magnetization may have contributions from inter-filament Bi-2212 contacts that induce currents circulating on lengths shorter than the 1 cm gauge length of the transport measurements \cite{angadi}. The field-increasing, field-decreasing behavior of transport \Jc{} is shown in Figure \ref{fig:hysteresis}. Strikingly, no sample exhibited significant \Jc{}(H) hysteresis, despite \Jc{}(H) values being much smaller for secondary HTs at 600 \degree{}C and higher. Transport \Jc{} hysteresis, present in even the best (and slightly overdoped) Bi-2223 tapes, is a signature of GB weak links, where trapped flux penetrating weakly coupled GBs effectively reduces the applied field on the return path of the field, increasing the inter-grain \Jc{} of the weak link network over which the supercurrent flows \cite{Dimos1988Orientation,Dyachenko1993Hysteresis,Evetts1988Relation}.  The magnitude of \Jc{} in Figure \ref{fig:main}f is typical for wires made with Nexans granulate powder \cite{Larbalestier2014Isotropic}, about 2600 Amm$^{-2}$ at 15 T, 4.2 K.  As discussed in the next section, more recent wires made with a finer, more uniform-sized powder have generated more than twice higher \Jc{}(4.2 K, 15 T) with a tighter biaxial texture and with the same lack of hysteretic \Jc{}(H) characteristics.

\begin{figure*}
	\includegraphics[width=2.0\columnwidth]{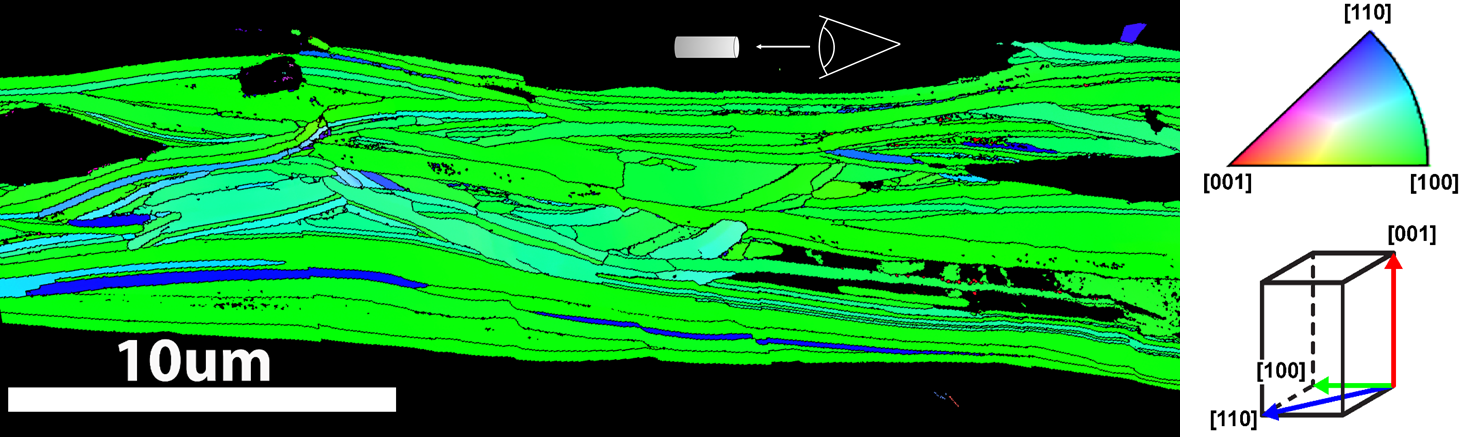}  
	\caption{\label{fig:EBSD} Inverse pole figure map (IPF) showing the grain orientations on a longitudinal cross section of an individual Bi-2212 filament in the highest \Jc{} sample in its initial fully-optimized post overpressure HT state. As shown in the IPF color legend, the grain orientations are represented by mixing green, blue, and red which correspond to Bi-2212’s [100], [110] and [001] crystallographic axes, respectively. The grain orientations are determined with the reference axis parallel to the filament direction. The dominance of the green color shows the strong [100] (\textit{a}-axis) grain alignment along the filament direction. As a result, most of the grain boundaries are also parallel to the filament direction and/or [100], indicating that the strongly \textit{a}-axis textured, large Bi-2212 grains form large area basal-plane-faced grain boundaries.}
\end{figure*}
\section{Discussion}
As noted earlier, extensive experiments and theoretical modeling of [001] tilt GBs between \textit{ab}-plane aligned grains, mostly for 123-YBCO but also for 2212 and 2223 bicrystals, all agree that weak link behavior starts at small misorientations of 5\degree{} or less, a major cause of weak link behavior being the lower hole and superfluid density of the GB compared to the grain. The depressed GB carrier density is often ascribed to ejection of O from the charge reservoir layers due to local GB or GB dislocation strains \cite{song2005electromagnetic,klie2005enhanced,schofield2004direct,gurevich1998current}. Even in the case of Ca-doped REBCO bicrystals, it is clear that the transport behavior of [001] tilt GBs improves near \Tc{} because \Tc{} depression caused by Ca addition is relieved by Ca desegregation away from the strongly-coupled channels between the GB dislocations \cite{Li2015Study,song2005electromagnetic,Kim2020Influence}. Thus, two observations of the present work are very significant:

\begin{enumerate}
  \item that a 15\degree{} FWHM biaxial texture clustered around [001], a misorientation well into the weak link regime of [001] tilt bicrystal experiments, does not produce observable weak link behavior, and
  \item that significantly underdoping Bi-2212 GBs in the current path does not produce observable weak link behavior. We therefore conclude that the quasi-biaxial texture of Bi-2212 intrinsically constitutes a strongly coupled current path, regardless of its oxygen doping state. This is a quite unexpected result, incompatible with the standard explanation of current transport in polycrystalline HTS cuprates that strong links and weak links coexist in the active current path and that the path to higher \Jc{} is to ameliorate the doping state of the weak-linked grain boundaries to one of higher carrier and superfluid density and consequently higher \Jc{}.
\end{enumerate}

There is a very important technological component to the present results.  The very best uniaxially textured ($\sim$15\degree{} FWHM) Bi-2223 tapes achieve \Jc{} (15 T,4.2 K) of $\sim$1100 Amm$^{-2}$ with a hysteretic field-increasing/field-decreasing weak-link characteristic, while these Bi-2212 round wires achieved $\sim$2600 Amm$^{-2}$ and recent wires made with finer and more uniform powder \cite{Jiang2019High} and similar processing can now achieve a \Jc{} (15 T,4.2 K) of $\sim$6500 Amm$^{-2}$, all without weak link hysteresis. Thus Bi-2223 seems to depend on weak-linked GBs for its high \Jc{} value while Bi-2212 with 6 times higher \Jc{} does not. Moreover, round wire Bi-2212 has a completely flexible architecture with respect to number of filaments, wire and filament diameter, \Ic{} rating and is also macroscopically isotropic and can be twisted to reduce AC loss \cite{YavuzACloss}. By contrast, Bi-2223 is anisotropic in shape and superconducting properties and made only in one size and one \Ic{} rating due to the complex reaction-deformation process that generates the uniaxial texture that maximizes its six-times lower \Jc{}. Yet, from an atomic structure point of view Bi-2223 and Bi-2212 are extremely close, sharing a common double Bi-O charge reservoir layer and differing principally by the fact that Bi-2223 has three CuO$_2$ layers and Bi-2212 just two. In both respects therefore it seems that round wire Bi-2212 breaks the paradigm that forces biaxially textured REBCO and uniaxially textured 2223 into tape geometries that reproduce the strong anisotropy of the native crystal structure and force expensive fabrication routes to ensure the best possible texture with minimum density of high angle grain boundaries.

This paradigm-breaking behavior in Bi-2212 can also be expressed in another way. Although its melt-textured biaxial texture is remarkable, its FWHM of 15\degree{} is not, being far into the weak-link regime associated with [001] tilt bicrystals \cite{Kametani2015Comparison}. What [001] tilt bicrystals of many compositions show is that there is a hybrid misorientation regime in the 5-10\degree{} range where the \Jc{} of parallel connected strongly coupled and weakly coupled portions of the GB is significantly degraded by withdrawing oxygen from the grain boundary \cite{Li2015Study,Kim2020Influence,Kim2020StrainDriven}. It is precisely this behavior \textbf{which is not seen here} when we underdope Bi-2212 round wires: while \Jc{} is reduced there is no sign of former well connected paths becoming weak linked.

We now should revisit the comparison of Bi-2212 and Bi-2223 in the context of the Railway Switch \cite{Hensel1993model} and Brick Wall \cite{Bulaevskii1992Model} models to explore how they might develop significantly different non-hysteretic and hysteretic \Jc{}(H) characteristics. It was clearly shown in Figure 5 of the comparative Bi-2212/Bi-2223 study of Kametani \textit{et al.} \cite{Kametani2015Comparison} that Bi-2223 grains have essentially random \textit{ab}-plane misorientations, while those of Bi-2212 are tightly clustered within about 15\degree{} of [100]. Such Inverse Pole Figures (IPFs) do not tell the whole story, however, as a further study by Kametani \textit{et al.} on two recent Bi-2223 tapes with \Jc{} values differing by 25\% shows \cite{kametani2019visualization}. As in reference \cite{Kametani2015Comparison}, many Railway Switch type SCTILT intersections in which small angle \textit{c}-axis tilt grain boundaries obliquely butt into the basal plane of another are present in these Bi-2223 tapes. Although not many SCTILT bicrystals have been studied, a few did fall into the angular range studied by Dimos \textit{et al.} \cite{Dimos1990Superconducting} where they were shown to be weak-linked for even small misorientations, consistent with the weak-link \Jc{}(H) hysteresis always measured in even the best Bi-2223 tapes. By contrast Bi-2212 grains tend to align their basal planes parallel to each other, often by plastic bending accommodation during the slow cooling from the melt stage of the heat treatment. The question of whether such basal-plane faced grain boundaries would be weak links has in fact been studied by sintering single crystals together \cite{Wang1994Electromagnetic,Li1997Superconducting,Zhu1998Structural,Li2000Electromagnetic} and measuring the in-plane and \textit{c}-axis transport \Ic{} properties. Very strikingly it was found that \Ic{} was independent of [001] twist misorientation at the boundary and that no distinction could be made for \Ic{} measured with voltage taps along the \textit{ab}-plane and across the grain boundary for \textit{c}-axis current flow. Thus there is some evidence that Bi-2212 [001] twist GBs connecting high aspect ratio grains, such as those found in textured tapes or in biaxially textured Bi-2212 round wires, can act as strongly coupled Brick Wall type connections limited by intra-grain vortex pinning.

The \textit{ab}-plane to \textit{c}-axis grain aspect ratio is the foundation of the Brick Wall model. In Bulaevskii’s model, the brick wall cartoon is of a standard brick with a 2:1 length to thickness ratio. But, as shown in the IPF image of Figure \ref{fig:EBSD}, Bi-2212 grains have a much more anisotropic shape, resulting in a length-thickness ratio of 40:1 or more. Figure \ref{fig:EBSD}, a longitudinal filament slice, clearly shows a strong [100] grain texture, in which the long dimension of aligned grains in the \textit{a}-axis direction and their propensity to smoothly align their basal planes together is apparent. Most of the grains are more than 40 $\mu$m long, much longer than the grain thickness (1 $\mu$m maximum), which creates large basal-plane contact areas suitable for \textit{c}-axis current flow around obstacles in the \textit{ab}-plane current path, even though the \textit{c}-axis \Jc{} is much lower than the \textit{ab}-plane \Jc{}. The strong \textit{a}-axis texture means these anisotropic Bi-2212 grains tend to stack parallel to each other, forming a large basal plane grain boundary area with only small misorientations that can favor \textit{c}-axis transport. Bi-2212 grains can plastically bend during their high-temperature, slow cooling step, so as to minimize the strain between slightly misaligned grains, partially relieving the \textit{c}-axis tilt misorientation and enhancing the basal-plane GB contact area. Thus, the dominant grain intersections in Bi-2212 are clearly the basal-plane faced OABTWIST GBs of Bulaevskii \textit{et al.} \cite{Bulaevskii1992Model}, despite the presence of some ECTILT or SCTILT grain boundaries that we expect from bicrystal studies to be weak-linked. The fact therefore that we do not observe weak-link hysteresis in \Jc{}(H) is consistent with the Brick Wall mode of supercurrent transport in Bi-2212, where the \Ic{} of this strongly coupled Brick Wall type current path is limited by intra-grain vortex pinning strength. We note that this strongly coupled current path is only a fraction of the superconducting volume, and we believe that advances in Bi-2212 RW \Jc{} come principally from increasing this fraction \cite{brown,ShaonDistribution}.

\section{Summary}
We have made a careful study of the influence of underdoping on the superconducting properties of optimized Bi-2212 round wire superconductors. Very surprisingly we find no evidence of weak link behavior, even in the most underdoped state when both grains and grain boundaries are strongly underdoped. Detailed EBSD study of the grain structure shows that most grain-to-grain contacts occur by basal-plane-faced grain boundaries that earlier bicrystal studies have shown not to be weak linked. We conclude that modern Bi-2212 wires exemplify the non-weak-linked “Brick Wall” model of supercurrent transport around obstacles in the current path, as opposed to the weak-linked current path common in optimized Bi-2223 tapes that has been attributed to “Railway Switch” transport. Moreover, round wire 2212 has recently doubled its \Jc{} values by using more uniform, finer particle powders. It thus seems that this method of developing Brick Wall current transport may still offer further opportunities to raise the critical current density of this most intriguing, round wire, macroscopically isotropic and multifilament cuprate superconductor.

\begin{acknowledgments}
The authors would like to thank Van Griffin and Ashleigh Francis for helping with and performing some of the transport \Ic{} and magnetization measurements, and Chiara Tarantini for discussions and suggestions. This work was performed at the National High Magnetic Field Laboratory, which is supported by the NSF under Award Numbers DMR-1157490 and DMR-1644779, and by the State of Florida. Primary support for this work was supplied by the US DOE Office of High Energy Physics under grant number DE-SC0010421.
\end{acknowledgments}


\bibliography{strongly_coupled_Bi2212}
\end{document}